\documentclass[aps, prl, reprint, lengthcheck, footinbib, showpacs, superscriptaddress]{revtex4-1}
\usepackage[T1]{fontenc}
\usepackage[english]{babel}
\usepackage[utf8]{inputenc}
\usepackage{textcomp}
\usepackage{amsmath}
\usepackage{braket}
\usepackage{graphicx}
\usepackage{epstopdf}
\usepackage[cmyk]{xcolor}
\usepackage[colorlinks, urlcolor=blue, citecolor=blue, linkcolor=blue, pdfstartview=FitH]{hyperref}
\usepackage[all]{hypcap}
\usepackage{times}

\begin{abstract}
The optical gyrotropy noise of a high-finesse semiconductor Bragg microcavity with an embedded quantum well (QW) is studied at different detunings of the photon mode and the QW exciton resonances. A strong suppression  of the noise magnitude for the photon mode frequencies lying above exciton resonances is found. We show that such a critical behavior of the observed optical noise power is specific of asymmetric Fabry-Perot resonators. As follows from our analysis, at a certain level of intracavity loss, the reflectivity of the asymmetric resonator vanishes, while the polarimetric sensitivity to the gyrotropy changes dramatically when moving across the critical point. The results of model calculations are in a good agreement with our experimental data on the spin noise in a single-quantum-well microcavity and are confirmed also by the spectra of the photo-induced Kerr rotation in the pump-probe experiments. 
\end{abstract}
\begin{document}

\title{Optics of spin-noise-induced gyrotropy of asymmetric microcavity }

\author{S.~V.~Poltavtsev}
\author{I.~I.~Ryzhov}

\author{R.~V.~Cherbunin}
\author{A.~V.~Mikhailov}

\author{N.~E.~Kopteva}

\author{G.~G.~Kozlov}

\affiliation{Spin-Optics laboratory, St.~Petersburg State University, 198504 St.~Petersburg, Russia}

\author{K.~V.~Kavokin}
\affiliation{Spin-Optics laboratory, St.~Petersburg State University, 198504 St.~Petersburg, Russia}
\affiliation{A.~F. Ioffe Physical-Technical Institute, Russian Academy of Sciences, 194021 St.~Petersburg, Russia}

\author{V.~S.~Zapasskii}
\affiliation{Spin-Optics laboratory, St.~Petersburg State University, 198504 St.~Petersburg, Russia}  


\author{P.~V.~Lagoudakis}
\affiliation{Department of Physics \& Astronomy, University of Southampton, Southampton SO17 1BJ, United Kingdom}

\author{A.~V.~Kavokin}
\affiliation{Spin-Optics laboratory, St.~Petersburg State University, 198504 St.~Petersburg, Russia}
\affiliation{Department of Physics \& Astronomy, University of Southampton, Southampton SO17 1BJ, United Kingdom}

\date{\today}
\maketitle 

\section{INTRODUCTION}

Optical resonators comprised of two plane-parallel mirrors (Fabry-Perot
cavities) play an important role in many optical applications and are studied
in great detail, presently \cite{B&W}. The growth of interest to this simple
resonant system in the last decade is related to advances in technology of
small-size optical resonators (microcavities) and possibility to place
structures with material resonances inside the cavity (quantum wells or
quantum dots), whose interaction with the optical mode of the microcavity
gives rise to a number of interesting optical effects \cite{Kavokin}.

Ability of the cavity to amplify various optical effects including the
magneto-optical Faraday and Kerr rotation (see, e.g., \cite{Zap0,Kav1,Kav2})
is highly valuable for the spin noise spectroscopy \cite{Zap,Zap1} gaining,
in the last years, especial popularity as a method of non-perturbative study
of magneto-spin properties of atomic and semiconductor systems \cite%
{Crooker,Os,Mitsui}. In the experiments of this kind (one of which will be
considered below), the above ability of the microcavity makes it possible to
observe the spin-induced gyrotropy noise of a thin inter-mirror gap, whose
detection without microcavity would have been strongly humpered.

The microcavities currently used in experimental studies are, most
frequently, asymmetric. A thorough analysis of their (unusual) polarization
properties is important. In this paper, we study the dependence of the 
\textit{optical spin noise} (OSN) spectra \cite{f1,Zap2} on the detuning of
the microcavity photon mode with respect to the exciton resonance \cite%
{Glazov}. Our experiments show that a shift of the photon mode frequency
below the exciton resonance leads to a sharp increase in the  sensitivity of
the polarisation-dependent microcavity reflectivity to fluctuations of the
gyrotropy, which cannot be simply ascribed to the finesse changes. We
attribute this critical behavior to specific phase characteristics of the
used asymmetric Bragg microcavity \cite{Salis,f2}. 

\section{EXPERIMENTAL}

\begin{figure}[tbp]
\includegraphics[width=.8\columnwidth,clip]{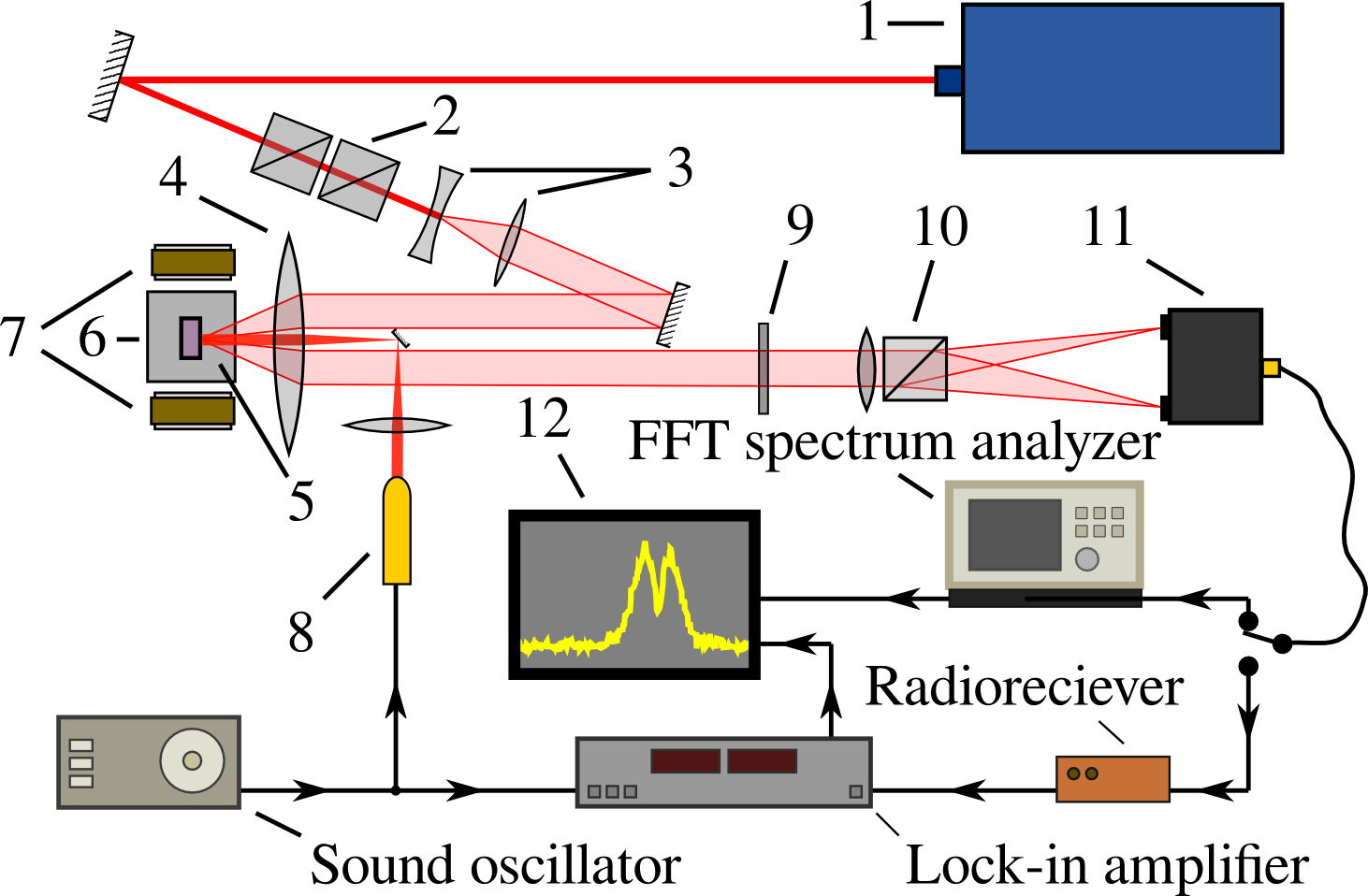}
\caption{ Schematic of the experimental setup. }
\label{fig_1}
\end{figure}

In this study, we used the sample described in \cite{Glazov}, representing
an asymmetric $\lambda$-microcavity (15 and 25 periods of GaAlAs/AlAs layers
in the top and bottom mirrors, respectively) with a GaAs single QW
(102/200/102 \hskip1mm \AA \hskip1mm AlAs/GaAs/AlAs) in the center of the
inter-mirror gap. In our experimental setup (Fig. \ref{fig_1}), a
monochromatic beam of tunable laser (1) transmitted through an attenuator
and linear polarizer (2), was focused by lens (4) onto the sample (5) placed
into a closed-cycle cryostat (6) cooling the sample down to $\sim$5 K. The
beam reflected from the sample was directed to a polarimetric detector
comprised of phase plate (9), polarization beamsplitter (Wollaston prism)
(10), and broadband ($\delta\nu$ = 200 MHz) balanced detector (11).

The noise signal from the sample, as a rule, was observed in the presence of
an additional illumination of the spot by a cw red light (photodoping) \cite%
{Glazov,Rapaport2001}, created by a laser diode (8) ($\sim $2 mW, $\lambda
\sim $630 nm). Fluctuations of spin polarization of the carriers
photo-generated in the cavity gave rise to fluctuations of the reflected
beam polarization and appearance of the noise signal at the output of the
photodetector (11). A typical experiment on spin noise spectroscopy \cite%
{Zap,Crooker,Os,Mitsui} implies studying the radio-frequency (RF) spectrum of
this signal as a function of the transverse magnetic field. This spectrum is
detected using a spectrum analyzer and, in the simplest case, has a
Lorentzian shape with its peak shifting linearly with magnetic field. 
Since the task of this paper was to study optical (rather than RF) noise
spectrum, the experiments were performed using a new scheme of the
measurements. The output of the detector was fed not to a spectrum analyzer,
but rather to a short-wavelength radioreceiver (resonant amplifier with an
output detector) tuned to the frequency $\omega $ = 35 MHz. With the aid of
a sound oscillator, the illumination was modulated at the frequency $\Omega $
= 60 Hz. This modulation produced synchronous modulation of the noise with
corresponding periodic signal at the output of the radioreceiver, with its
amplitude proportional to that of the noise at the frequency 35 MHz. With
the use of a lock-in amplifier (Fig. \ref{fig_1}), we recorded the optical
spin noise (OSN) spectrum, i.e., dependence of the above noise signal versus
the wavelength of tunable laser (1). The experiments were performed at zero
magnetic field, because the noise amplitude at the frequency 35 MHz, in this
case, was high enough for its detection (due to the large width of the spin
noise spectrum \cite{Glazov}).

\begin{figure}[tbp]
\includegraphics[width=.8\columnwidth,clip]{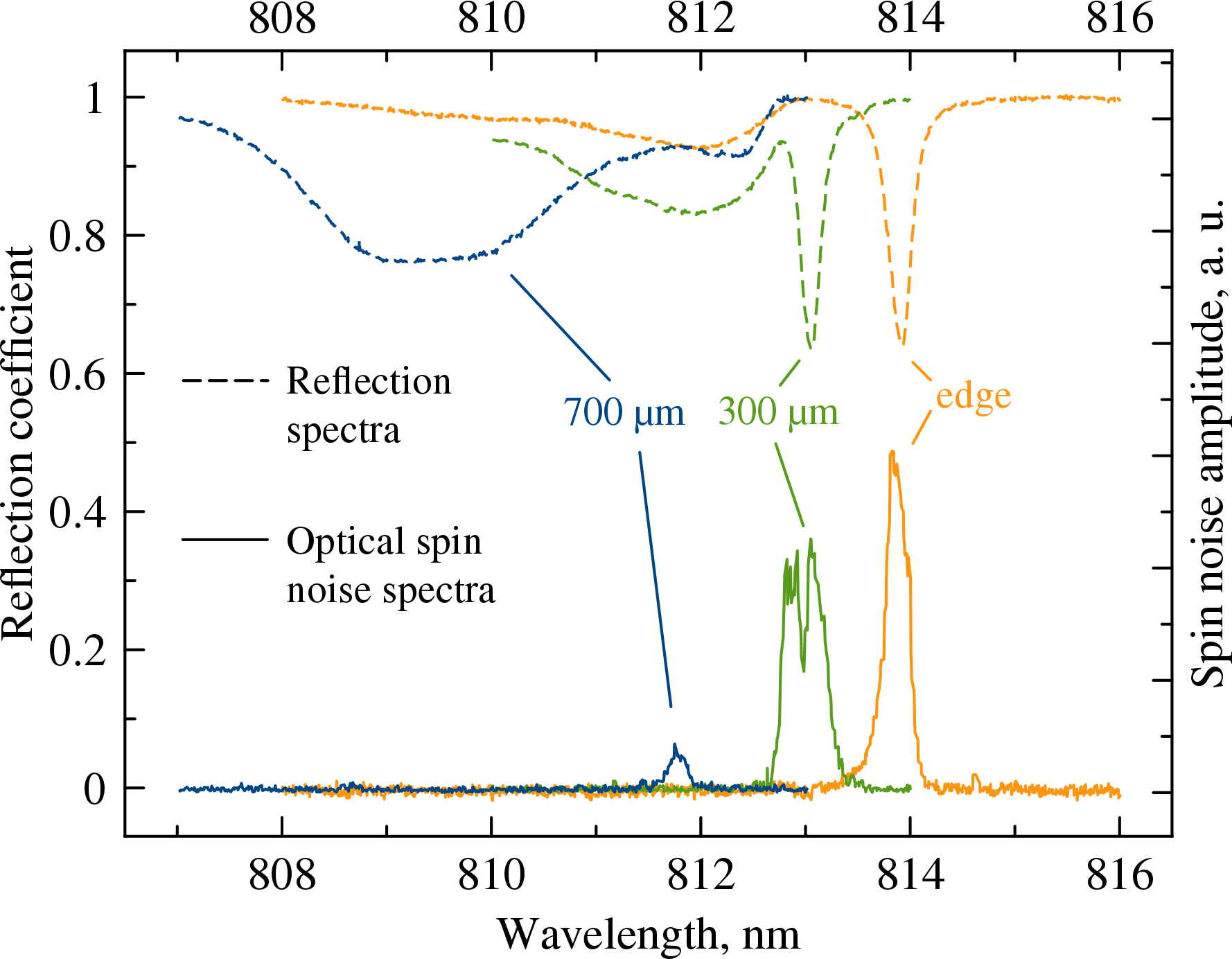}
\caption{ OSN spectra at different relative spectral position of the photon
mode and exciton resonances of the QW (figures in $\protect\mu $m indicate
distance of the light spot from the edge of the sample). The photon mode
frequency varies with position of the probe beam on the sample. Edge -- the
photon mode lies below the exciton resonance of the QW, and the OSN spectrum
has a monomodal shape. In the vicinity of anticrossing of the polaritonic
branches, the spectrum becomes bimodal (300 $\protect\mu $m), then it
decreases in amplitude (700 $\protect\mu $m) and becomes unobservable at
the photon mode frequency above the exciton resonances of the QW. The upper
plots show the reflectivity spectra in corresponding points of the sample.}
\label{fig_2}
\end{figure}

Figure \ref{fig_2} shows the OSN spectra obtained using the above method at
different detunings of the photon mode of the cavity and the exciton
resonance of the QW. Due to spatial gradient of the inter-mirror gap
thickness, the photon mode frequency could be shifted by moving the light
spot of the probe beam at the surface of the microcavity. It is seen from
the figure that the evolution of the OSN spectrum with the photone mode
frequency demonstrates three distinct regimes: (i) For the photon mode
frequencies lying below the exciton resonance (the region of negative
detuning), the OSN spectrum has a shape of a narrow peak with its spectral
position coincident with that of the photon mode. In the spectral region of
the QW exciton resonances (which are already noticeable in the reflectivity
spectrum, see Fig. \ref{fig_2}, upper curves), the OSN spectrum amplitude is
vanishingly small. (ii) As the photon mode frequency approaches those of the
exciton resonances (the region of anti-crossing of the polariton branches),
the amplitude of the OSN spectrum decreases, while its shape deviates from
monomodal, with arising two (or three) peaks. Spectral position of the peaks
is close to position of the photon mode of the microcavity. In the spectral
region of the QW exciton resonances, amplitude of the OSN spectrum remains
small. (iii) When the photon mode frequency starts to exceed those of the
exciton resonances (the region of positive detuning), the OSN spectrum
amplitude falls down to unobservable level. The amplitude of the
reflectivity spectrum for this region of the sample is about 30 - 50 $\%$
lower than for the region with negative detuning. Note that the width of the
photon resonance, in the reflectivity spectrum, for the region with a
positive detuning is by a factor of 3 - 4 larger than for the region with a
negative detuning (the reflectivity spectra of different regions of the
sample are given in \cite{Glazov}). 
The sharp decrease of the OSN amplitude when passing from negative to
positive detuning, pointed out in item (iii) is inconsistent with the
decrease of the microcavity finesse and looks surprising. The main goal of
this paper is to interpret this unusual behavior of the Kerr-rotation noise.

\section{DISCUSSION}

In what follows, we assume, for simplicity, that the main role in the
formation of the OSN spectra is played by fluctuations of gyrotropy of the
inter-mirror gap as a whole, rather than only of the QW excitons. Since the
transparent cavity material does not have any spectral features in the
studied wavelength range, we may neglect, in our qualitative treatment, the
spectral dependence of the gyrotropy $\delta g$ responsible for the observed
noise signal. This assumption is not indisputable, but the qualitative
agreement of the model calculations with the experimental data evidence in
its favor.

Formation of the polarimetric signal detected in our experiment can be
presented in the following way. The linearly polarized beam incident upon
the sample is a superposition of two circularly polarized components $\sigma
_{+}$ and $\sigma _{-}$of equal magnitudes. In the presence of gyrotropy in
the microcavity, positions of resonant features of its reflectivity spectrum
for the $\sigma _{+}$ and $\sigma _{-}$ components of the incident light
appear to be shifted with respect to each other by the quantity proportional
to this gyrotropy. Thus, at $\delta g\neq 0$, the $\sigma _{+}$ and $\sigma
_{-}$ components of the incident beam will experience, upon reflection,
different phase shifts and different amplitude changes. The former
corresponds to polarization plane rotation of the reflected beam, while the
latter manifests the appearance of its ellipticity. Both types of the
polarization changes in the reflected beam can be detected independently. In
particular, if the phase plate (9) (Fig. \ref{fig_1}) is half-wave
(quarter-wave) and is oriented by its axis at the angle $\pi /8$ ($\pi /4$)
with respect to the vertical, the output signal of the balanced detector
(11) will be proportional to the polarization plane rotation angle
(ellipticity) of the reflected beam. 
Note that, at a fixed frequency shift of the phase characteristics of reflectivity for the $\sigma_+$ and $\sigma_-$ components of the incident light, the changes in the reflected beam polarization will increase with increasing steepness of frequency dependence of these characteristics. 
\begin{figure}[tbp]
\includegraphics[width=.8\columnwidth,clip]{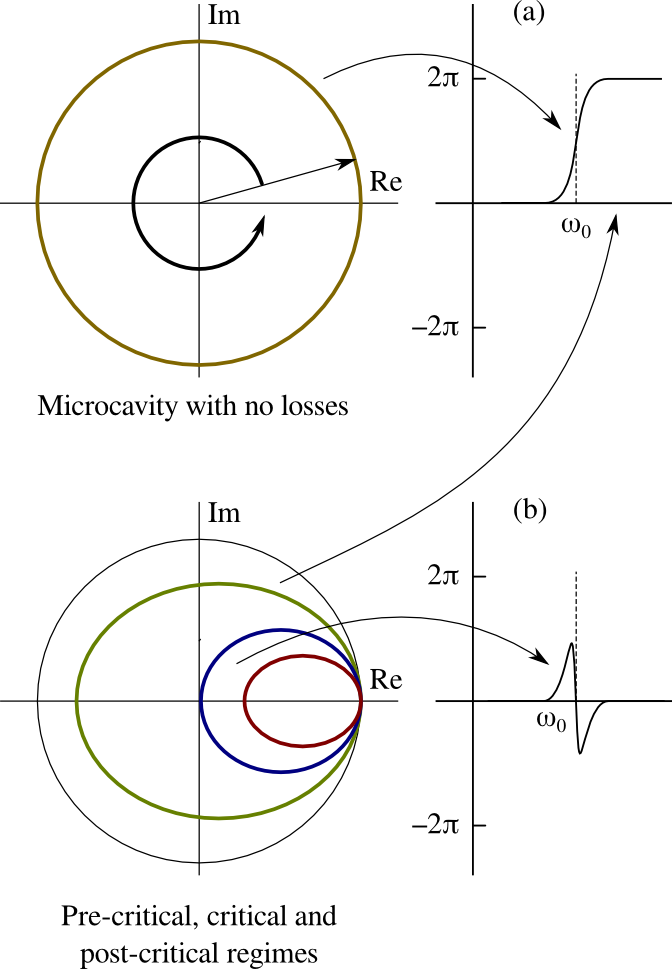}
\caption{Dynamics of the complex reflectivity vector of the asymmetric
cavity upon passing through the resonance. (a) -- ideal (ultimate)
asymmetric cavity, (b) -- asymmetric cavity at different absorption in the
inter-mirror gap. The right plots show phase of the cavity reflection
coefficient in the pre- and post-critical regime. }
\label{fig_3}
\end{figure}
Thus, for interpretation of the above experiment, we have to analyze
behavior of the complex reflection coefficient of the microcavity with
variation of optical parameters of the cavity. This kind of analysis has been
performed for interpretation of a number of other experiments (see, e.g. 
\cite{Salis,Kav1,Kav2}) and for development of interferometric devices with
specific dispersion chahracteristics \cite{Tr1,Tr2}. The analysis presented
below is aimed at interpretation of our particular spin-noise-related
experiment.

As we have already mentioned, a specific feature of the microcavity used in
our experiment is its asymmetry. The complex reflection coefficient of such
a microcavity reveals a specific critical behavior, which can be
qualitatively explained as follows.

Consider an ultimately asymmetric cavity, with its bottom mirror having
reflectivity exactly equal to unity, while reflectivity of the top mirror is
slightly smaller. 
 If the absorption in the inter-mirror gap and top mirror is absent, then the reflectivity module of such a cavity is unity at any frequency of the incident light.
 As can be shown by direct calculations,
the phase of the reflection coefficient is zero (2$\pi $) at frequencies
much lower (higher) than the resonant one, and it rapidly changes from zero
to 2$\pi $ in the vicinity of the resonance whose spectral width is
controlled by the cavity finesse, which, in turn, depends on the top mirror
reflectivity. Thus, upon variation of the incident light frequency, the
complex reflection coefficient can be represented by a  vector which
circumvents a full circle moving over it counter-clockwise from the real
unity (Fig. \ref{fig_3},a).

Let us now assume that the cavity is filled by an absorbing material\textbf{.%
} In this case, at the frequencies far from the resonance, the reflectivity
of the cavity will be controlled by reflectivity of the top mirror and,
since we assume this mirror to be highly reflecting, will be, as before,
close to unity. In the region of the resonance, the reflectivity will reveal
a small dip connected with absorption in the inter-mirror gap. As a result,
dynamics of the complex reflection coefficient, upon passing the resonance,
will differ from that for the above ultimate cavity: the corresponding
complex vector will now circumvent a certain oval lying inside the above
unity circle (Fig. \ref{fig_3},b). If the absorption in the cavity is not
high, this oval will not much differ from the unity circle, and the
increment of the phase of the reflection coefficient, upon passing through
the resonance, as before, will be equal to 2$\pi $. With further growth of
the absorption, the oval will contract to the point of real unity, because,
even at very high absorption in the cavity, the reflectivity far away from
the resonance will be determined by the top highly reflecting mirror. It
follows herefrom that, at some critical value of the absorption, the oval
will pass through the coordinate origin and will appear to be entirely at
the right side of this point (Fig. \ref{fig_3},b). It is seen from Fig. \ref%
{fig_3},b that, after that, the total increment of the phase of the
reflection coefficient, upon passing through the resonance, turns into zero.
This means that the phase, which changes monotonically, in the pre-critical
regime, from 0 to $2\pi $ (Fig.\ref{fig_3}, right plot), will experience, in
the post-critical regime, a relatively small ($< \pi /2$)
sign-alternating change (Fig. \ref{fig_3},b, right plot).

As was already noted, the gyrotropy of the inter-mirror gap gives rise to
mutual shift of the phase of the reflection coefficients for the $\sigma _{+}
$ and $\sigma _{-}$ components of the incident light, with magnitude of the
polarimetric signal being higher for steeper phase characteristics. Since
the phase of the reflection coefficient of the pre-critical cavity (Fig. \ref%
{fig_3},a) is much steeper than of the post-critical one (Fig. \ref{fig_3}%
,b), one can expect that the polarimetric signal of the pre-critical cavity
will be much larger than that of the post-critical one. Besides, the
transition of the cavity through the critical point should be accompanied by
a change of optical spectrum of the polarimetric signal, because this
spectrum (at least, qualitatively) is controlled by the frequency derivative
of the phase of the reflection coefficient. In the pre-critical regime, the
phase is monotonic, which leads to monomodal shape of optical spectrum of
the polarimetric signal. In the post-critical regime, the non-monotonic
dependence of the phase over the detuning should result in the
sign-alternating shape of spectrum of the polarimetric signal.

\begin{figure}[tbp]
\includegraphics[width=.8\columnwidth,clip]{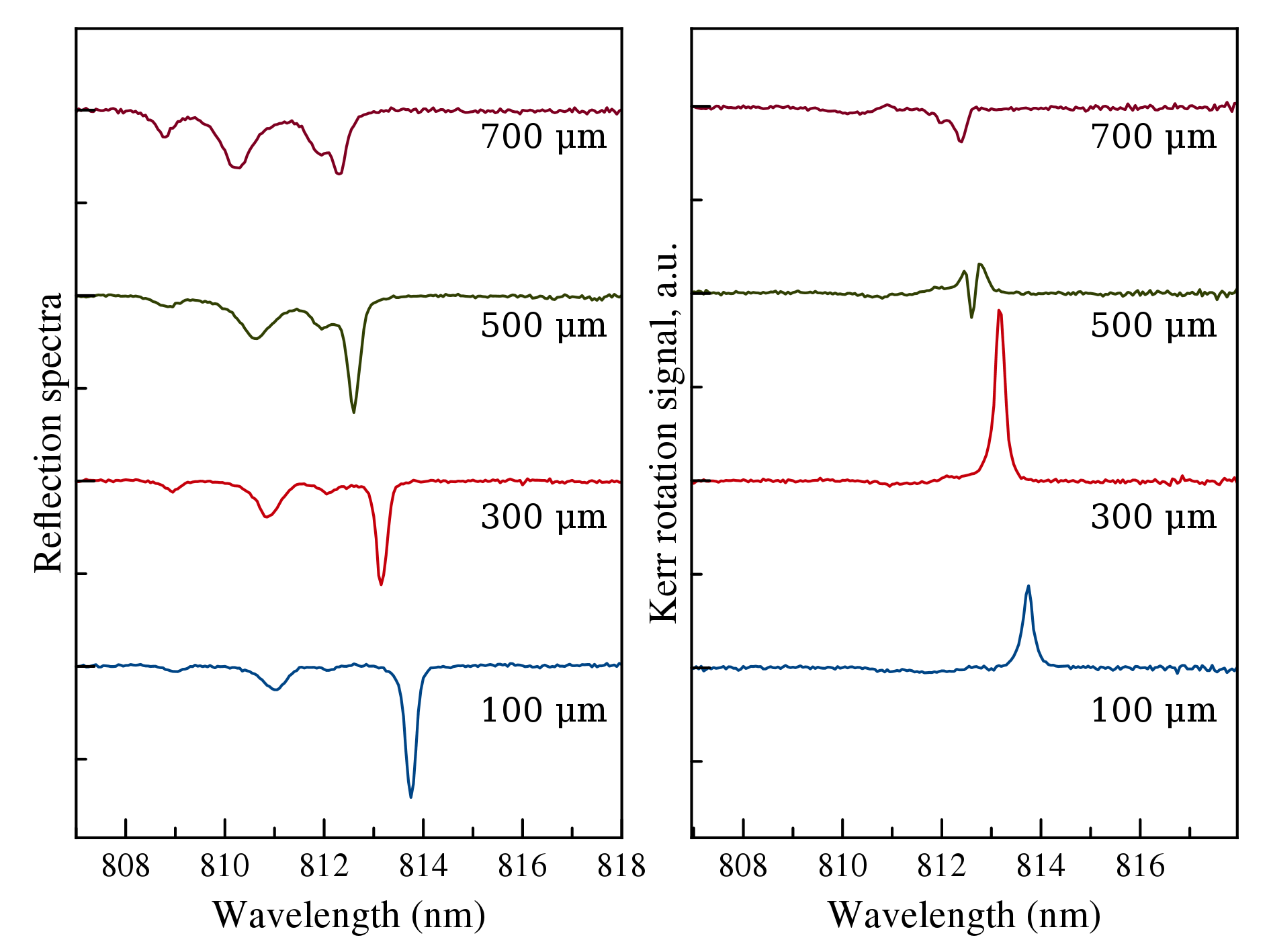}
\caption{Right panel: evolution of shape of the Kerr rotation spectra upon
shifting of the photon mode of the cavity towards shorter wavelengths. The
measurements were performed in the pump-probe configuration. The gyrotropy
of the sample was induced by a high-power pulse of circularly polarized
light and was detected by a weak probe light, whose Kerr rotation was
measured as a function of its wavelength. The figure shows the results of
the measurements for different points on the sample: from point 100 
$\mu $m to point 700 $\protect\mu $m, position of the photon mode shifts
towards shorter wavelengths. Left panel: the corresponding reflection
spectra.}
\label{fig_5}
\end{figure}

The property of the asymmetric cavity described above seems to be important
for the following reasons. First, the Bragg microcavity used in our
experiments is essentially asymmetric (15 periods in the top mirror and 25
periods in the bottom one), and the above reasoning is applicable to it.
Second, when spectral position of the photon mode moves towards shorter
wavelengths, absorption in the microcavity increases: this is confirmed by
more than three-fold broadening of the photon-mode-related line in the
reflectivity spectrum. And, finally, third, the observed evolution of the
optical spectra of Kerr rotation from monomodal to sigh-alternating (Fig. 
\ref{fig_5}) also correlates with passing of the used asymmetric
microcavity through the critical point.

For all the above reasons, the sharp fall in the OSN amplitude observed in
our experiment can be attributed to passing through the above critical point
of the microcavity. The direct model calculations described in the next
section support this supposition. 

\section{MODEL CALCULATIONS}

Since the angle of incidence of the laser beam, in our experiments, was
small ($\sim $ 5$^{o}$), we will assume the incidence to be normal, in the
model. Let us denote by $r$ the complex reflection coefficient of the
microcavity in the absence of gyrotropy in the inter-mirror gap. With
appearance of the gyrotropy, the reflection coefficients $r_{\pm }$ for the $%
\sigma _{\pm }$ polarizations become different, and $r_{\pm }=r\pm \delta r$%
. The calculations show that, if the beam incident on the microcavity is
polarized vertically, the output signal $S$ of the balanced detector (11)
(Fig. \ref{fig_1}) is given by the relationship 
\begin{equation}
S=2|r|^{2}\bigg [\cos ^{2}2\phi +\cos \delta \hskip1mm\sin ^{2}2\phi \bigg ]+
\label{15}
\end{equation}%
\begin{equation*}
+2\sin [4\phi ]\hbox{ Im }\lbrack \delta rr^{\ast }][1-\cos \delta ]+4%
\hbox {
Re }\lbrack \delta rr^{\ast }]\hskip1mm\sin \delta \hskip1mm\sin [2\phi ],
\end{equation*}%
where $\delta $ is the phase shift introduced by plate (9) and $\phi $ is
the angle of tilt of its axis with respect to the vertical. By choosing the
angle $\phi $, the contribution to the signal independent of $\delta r$
(first line in Eq. (\ref{15})) can be vanished. For detecting fluctuations
of the Kerr rotation (ellipticity) of the reflected beam, the phase plate
(9) is chosen half-wave (quarter-wave). In this case, the detected signal
can be presented in the form $S=4\hbox { Im }\lbrack \delta rr^{\ast }]$ $%
(S=4\hbox { Re }\lbrack \delta rr^{\ast }])$.

Calculation of the OSN spectrum was started from calculation of the
reflection coefficient $r$ entering into (\ref{15}) and was performed using
the standard transfer matrix technique \cite{B&W}, with the QW
susceptibility $\varepsilon (\omega )$ taken in the model of independent
electron-hole pairs: 
\begin{equation}
\varepsilon (\omega )=\varepsilon _{b}+C\ln (\omega _{0}-\omega +\imath
\gamma ),  \label{2}
\end{equation}%
where $\omega _{0}$ = 1.516 eV is the lowest frequency of transition between
the 2D electron and hole sub-bands of the QW, $\gamma $ = 0.001 eV is the
damping constant, and $\omega $ is the optical frequency. The parameters $%
\varepsilon_b $ = 34 and $C$ = 3 were chosen to obtain a qualitative agreement
between the experimental and calculated reflectivity spectra \cite{f3}

\begin{figure}[tbp]
\includegraphics[width=1.\columnwidth,clip]{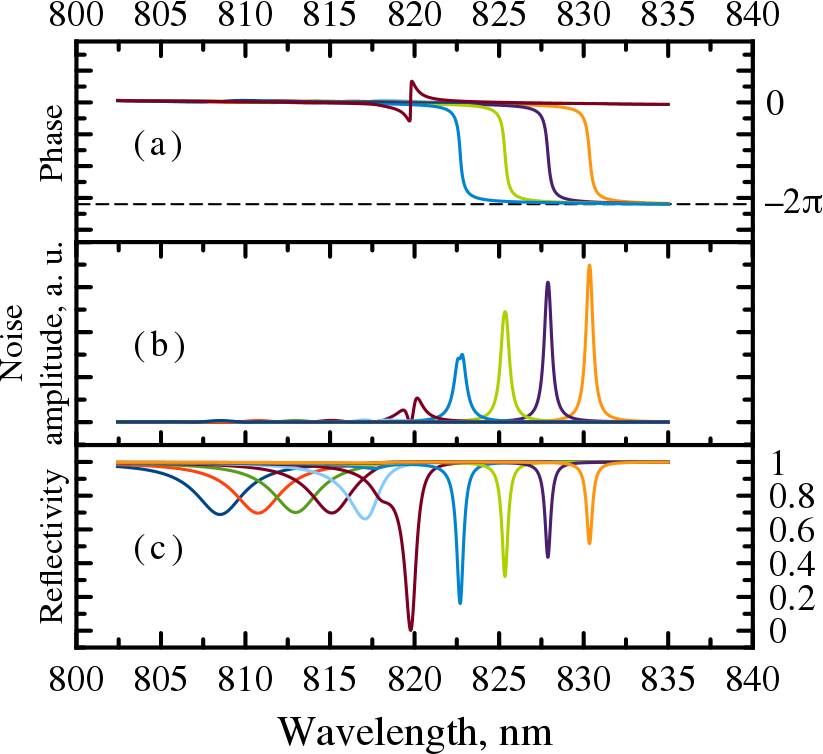}
\caption{The calculated phase of the reflection coefficient (a), OSN spectra
(b), and reflectivity spectra (c) of the asymmetric microcavity with QW at
different spectral position of the photon mode. The QW permittivity was
taken in the form $\protect\varepsilon (\protect\omega )=\protect\varepsilon %
_{b}+C\ln (\protect\omega _{0}-\protect\omega +\imath \protect\gamma )$,
with spectral position $\protect\omega _{0}$ corresponding to the center of
the shown spectral range.}
\label{fig_4}
\end{figure}

At the next step, the permittivity of the inter-mirror gap $\varepsilon $ 
 was privided 
with a small increment $\varepsilon \rightarrow \varepsilon +\delta
\varepsilon $ ($\delta \varepsilon /\varepsilon \ll 1$) corresponding to
noise fluctuation of the gyrotropy $\delta g$, and the corresponding change
of the reflection coefficient $\delta r$ was calculated. After that, the OSN
spectrum of the Kerr rotation (ellipticity) was calculated using Eq. (\ref%
{15}) at $\phi $ = $\pi /8,\delta =\pi $ ($\phi =\pi /4,\delta =\pi /2$) in
the range of the optical frequencies $\omega /\omega _{0}\in \lbrack
0.98,1.02]$. The results of the calculations are given in Fig. \ref{fig_4},
which shows a family of 
the phases of reflection coefficients
 (a), the OSN spectra of the Kerr rotation 
 (b), and the reflectivity spectra (c) at
different frequencies of the photon mode. The photon mode was shifted by
changing the cavity width within the range of 6\%. It is seen from Fig. \ref%
{fig_4} that, at the frequency of the photon mode below $\omega _{0}$, the
resonant absorption of the microcavity is small, and it lies in the
pre-critical regime. This is indicated by monotonic behaviour of the phase
(Fig. \ref{fig_4},a) varying within the interval from 0 to $2\pi $. In this
case, the OSN spectra have a monomodal shape and large amplitude (Fig. \ref%
{fig_4},b). As the frequency of the photon mode approaches $\omega _{0}$,
the absorption in the microcavity increases, the cavity passes through the
critical point, and the phase of the reflection coefficient becomes
sign-alternating (Fig. \ref{fig_4},a). It is also seen from Fig. \ref{fig_4}%
,b, that the shape of the OSN spectrum becomes, in this case, bimodal, and
its amplitude decreases. After passing through the critical point, the
reflectivity spectrum broadens by a factor of 3-4 and slightly drops in
amplitude, while the amplitude of the noise spectrum decreases dramatically.
Thus, the evolution of the calculated OSN spectra is characterized by three
regimes listed in the end of Section "Experimental". Qualitative agreement
between the behavior of the calculated and experimental OSN spectra confirms
the validity of our model. 

\section{CONCLUSIONS}

We have studied the behavior of the Kerr rotation noise spectrum of light
reflected from an asymmetric Bragg $\lambda $-microcavity with an embedded
GaAs quantum well for different detunings between the photon mode and
exciton resonances. The Kerr rotation noise has been found to
exibit an unusual critical dependence on the detuning\textbf{.} At negative
detunings, the OSN spectrum shows a relatively high amplitude and monomodal
shape. When passing to positive detuning, the shape of the spectrum is
getting more complicated, and the amplitude of the signal sharply drops. The
effect is explained by the fact that the phase of the reflection coefficient
of the used asymmetric cavity, when passing to positive detuning, shows
critical change of its steepness. The results of this paper show that the
cavity-enhancement of the Kerr (or Faraday) rotation in an asimmetric
microcavity may strongly depend on its phase characteristic. This critical
dependence of convertion of intra-cavity girotropy to the polarisation-plane
rotation may be highly important for experimental studies of spin noise and
other effects of cavity enhanced optical anisotropy.

\section{Acknowledgements}

The financial support from the Russian Ministry of Education and Science
(contract No.11.G34.31.0067 with SPbSU and leading scientist A. V. Kavokin)
is acknowledged. This work has been partially funded by Skolkovo Institute
of Science and Technology (Skoltech) in the framework of the SkolTech/MIT
Initiative. It was carried out using the equipment of SPbU Resource Center
"Nanophotonics" (photon.spbu.ru). 


\begin{thebibliography}{99}
\bibitem{f1} Here, we mean dependence of the noise power on wavelength of a
quasi-monochromatic laser beam, whose polarization noise is detected.
\bibitem{f2} Under asymmetry we understand difference between the number of
pairs in the top and bottom Bragg mirrors.
\bibitem{f3} Susceptibility of a real QW contains along with (\ref{2}), a
resonant excitonic contribution. However, the effect of this contribution,
at large positive and negative detunings is small, and its account, as show
the calculations, does not change qualitatively the ON spectrum.
\end{thebibliography}
\end{document}